\documentclass[aps,prx,twocolumn,superscriptaddress,showpacs,floatfix,longbibliography]{revtex4-2}
\usepackage{amsmath,amssymb,amsfonts,float,graphics,epsfig,epstopdf,color,verbatim,tabularx,bm,multirow,appendix,hyperref}
\usepackage{lmodern}
\usepackage{color}
\usepackage{bm}

\DeclareMathOperator{\Tr}{Tr}

\def\bk{{\mathbf{k}}}
\def\bK{{\mathbf{K}}}

\def\br{{\mathbf{r}}}

\def\bq{{\mathbf{q}}}

\def\bG{{\mathbf{G}}}

\begin{document}
	
\title {Momentum space quantum Monte Carlo on twisted bilayer Graphene}
\author{Xu Zhang}\thanks{These two authors contributed equally}
\affiliation{Department of Physics and HKU-UCAS Joint Institute of Theoretical and Computational Physics, The University of Hong Kong, Pokfulam Road, Hong Kong SAR, China}
\author{Gaopei Pan}\thanks{These two authors contributed equally}
\affiliation{Beijing National Laboratory for Condensed Matter Physics and Institute of Physics, Chinese Academy of Sciences, Beijing 100190, China}
\affiliation{School of Physical Sciences, University of Chinese Academy of Sciences, Beijing 100190, China}
\author{Yi Zhang}
\affiliation{Kavli Institute for Theoretical Sciences, University of Chinese Academy of Sciences, Beijing, 100190, China}
\author{Jian Kang}
%\email[]{jkang@suda.edu.cn}
\affiliation{School of Physical Science and Technology and Institute for Advanced Study, Soochow University, Suzhou, 215006, China}
\author{Zi Yang Meng}
\email[]{zymeng@hku.hk}
\affiliation{Department of Physics and HKU-UCAS Joint Institute of Theoretical and Computational Physics, The University of Hong Kong, Pokfulam Road, Hong Kong SAR, China}
\affiliation{Beijing National Laboratory for Condensed Matter Physics and Institute of Physics, Chinese Academy of Sciences, Beijing 100190, China}

\date{\today}
		
\begin{abstract}
We report an implementation of the momentum space quantum Monte Carlo (QMC) method on the interaction model for the twisted bilayer graphene (TBG). The long-range Coulomb repulsion is treated exactly with the flat bands, spin and valley degrees of freedom of electrons taking into account. We prove the absence of the minus sign problem for QMC simulation when either the two valley or the two spin degrees of freedom are considered. By taking the realistic parameters of the twist angle and interlayer tunnelings into the simulation, we benchmark the QMC data with the exact band gap obtained at the chiral limit, to reveal the insulating ground states at the charge neutrality point (CNP). Then, with the exact Green's functions from QMC, we perform stochastic analytic continuation to obtain the first set of single-particle spectral function for the TBG model at CNP. Our momentum space QMC scheme therefore offers the controlled computation pathway for systematic investigation of the electronic states in realistic TBG model at various electron fillings. 
\end{abstract}

\maketitle

{\it Introduction}\,---\, Twisted bilayer graphene (TBG) and other moir\'e systems have attracted great theoretical~\cite{Trambly2010,Trambly2012,Bistritzer12233,ROZHKOV20161,Santos2007,Santos2012} and experimental~\cite{cao2018unconventional,cao2018correlated,chen2020tunable,kerelsky2019maximized,PhysRevLett.123.046601,lu2019superconductors,xie2019spectroscopic,Shen2020DTBG,Nuckolls_2020,pierce2021unconventional,moriyama2019,rozen2020entropic,liu2020spectroscopy,Shen_2021} interests in the condensed matter and 2D quantum material communities. As experiments have discovered the correlated insulating phases at various integer fillings and the proximite superconductivity (SC) phases, a key question arises is how to model and understand the properties of the insulating phases, both for their own sake and that could eventually provide a clue for the understanding of the mechanism of the superconductivity in TBG systems. 
	
Many experimental and theoretical works have indicated the interplay between the nontrivial topology and strong interaction as the essential ingredients for the understanding of the electronic correlations in such materials, therefore pointing out a proper model for TBG system shall be significantly different from the typical Hubbard-type Hamiltonian with on-site interactions~\cite{HCPo2018Fragile,HCPo2019,Bultinck2020PRL,PhysRevX.8.031089,PhysRevLett.122.106405,PhysRevB.98.045103,Kang2018,Koshino2018,Vladimir2019}. However, the nature of the insulating states discovered in the material is still under debate. On the one hand, the analytical and the Hartree-Fock calculations at various integer fillings have found the quantum anomalous hall (QAH) and the intervalley coherent (IVC) states as the ground states without breaking the translation symmetry, suggesting that the physics is similar to the quantum Hall ferromagnetism at the lowest Landau level (LLL)~\cite{YiZhang2020,Bultinck2020,hejazi2020hybrid,PhysRevLett.124.097601,JPLiu2019Pseudo,liu2020anomalous,JPLiu2021,PhysRevB.102.045107,SLiu2021,carr2019minimal,YHKwan2021}. Such similarity also led to the proposal of the skyrmion SC for the mechanism of SC discovered near the insulating phases at $\nu = \pm 2$~\cite{Chatterjee2020,Khalaf2021}. On the other hand, recent numerical calculations based on the density matrix renormalization group (DMRG)~\cite{JKang2020DMRG,PhysRevB.102.205111,PhysRevB.102.155429,Chatterjee2020} and exact diagonalization (ED)~\cite{xie2020tbg6,PhysRevB.98.081102,PhysRevB.98.075154,Potasz2021} have discovered a larger manifold of nearly degenerate states strongly competing with each other even in the strong coupling regime, hinting that such systems could contain much more complicated physics than that of the LLL. As a consequence, there is a crying need for applying more extensive numerical methodology that can unbiasedly solve larger system sizes to fully settle the mechanism of the insulating phases at various integer fillings.

But this is by no means an easy task. In TBG each moir\'e superlattice unit cell contains more than $10^4$ carbon atoms(in the case close to the first magic angle), resulting in the same number of bands in the moir\'e Brillouin zone (mBZ). It is impossible to include such a huge number of bands in any realistic calculations for strongly correlated electrons. Fortunately, all experiments have found that the correlated physics emerges only when the chemical potential lies inside the flat bands~\cite{cao2018unconventional,cao2018correlated,chen2020tunable} and the large band gap that separates the flat bands and remote bands allows one to focus on the flat bands only to study the electronic correlations\cite{xie2019spectroscopic}. While the parameters of the Hamiltonian has been changed by integrating out the states on the remote bands in the presence of the Coulomb interactions, the effective Hamiltonian is still given by the Bistritzer-MacDonald (BM) model in momentum space~\cite{JKang2020RG} with the projected Coulomb interactions onto the flat bands, that has been significantly simplified (yet still difficult) for realistic analytical and numerical calculations~\cite{Kang2019,JKang2020DMRG,song2020tbg2,bernevig2020tbg3,lian2020tbg4,bernevig2020tbg5,xie2020tbg6,Bultinck2020,JPLiu2021,Alavirad2020}

In light of the situation, the large-scale quantum Monte Carlo (QMC) method presents itself as the ideal choice of method to solve these TBG models at integer fillings. QMC solves the correlated electron lattice models in path-integral such that both static and dynamic properties, at finite temperature and ground state, can be obtained in unbiased manner with only statistical errors. The extrapolation to the thermodynamic limit is also possible, when the computation complexity increases polynomially with the system sizes, i.e. absence of minus sign problem. Many important features of correlated electron system such as the antiferromagnetic Mott insulator in square lattice Hubbard model~\cite{Hirsch1985}, non-Fermi liquid at quantum critical points~\cite{XYXu2017,ZHLiuPNAS2019}, to name a few, have been discovered from QMC simulations. In case of TBG, by now there have been few QMC simulations in real-space lattice model with extended interactions where interaction-driven topological state, IVC and translational symmetry-breaking insulators are found~\cite{PhysRevLett.123.157601,PhysRevX.11.011014,Liao_2021,PhysRevB.98.121406,HUANG2019310}. But generic and systematic QMC analsis for BM-type models with flat bands, spin and valley degrees of freedom and in particular, the long-range Coulomb interaction to be fully respected in momentum space, is still missing. 

This is the knowledge gap we want to fill in. In this work, we develop a momentum space QMC method~\cite{ZHLiuEMUS2019,ZHLiuPNAS2019,ZJWang2021,Ippoliti2018} for the aforementioned TBG models. We first prove  the absence of the minus sign problem for QMC simulation at integer fillings when either the two valley or the two spin degrees of freedom are considered. Then, by taking the realistic parameters of the twist angle and interlayer tunnelings into account, we benchmark the QMC data with the exact band gap obtained with a $6\times6$ momentum mesh in the mBZ, to reveal the insulating ground states at the charge neutrality point (CNP). Finally, by combining the QMC simulation with the stochastic analytic continuation, we obtain the first set of single-particle spectra at chiral limit and realistic parameter at CNP. Our momentum space QMC scheme therefore offers the controlled computation pathway for systematic investigation of the electronic states in realistic TBG model at various electron fillings. 

%specially considering that such a mommentum-space QMC has already been applied to other physical problems and once succeeded here, one can foresee the full solution of the TBG interacting models with both static and dynamic properties under control.

%At certain limits (chiral limit for example) the ground state and excitation gaps of these models can be obtained with certainty~\cite{Kang2019,lian2020tbg4,bernevig2020tbg5}. However, generically the model with realistic parameters (small twist angle, interlayer tunnelings, Coulomb interaction, etc) cannot be solved due to its strong coupling nature, even at integer fillings. 
%There are many attempts with Hatree-Fock type of mean-field analysis~\cite{YiZhang2020,Bultinck2020,hejazi2020hybrid,PhysRevLett.124.097601,JPLiu2021,PhysRevB.102.045107,SLiu2021,carr2019minimal,Vahedi2021}, and There are also DMRG attempts in both real-~\cite{BBChen2020} and momentum-space~\cite{JKang2020DMRG,PhysRevB.102.205111,PhysRevB.102.155429,chatterjee2020skyrmion} and RG analysis of the renormalization of the narrow bands from Coulomb interaction~\cite{JKang2020RG}, and exct diagonalization (ED) on relatively small momentum-mesh in moir\'e Brillouin zone (mBZ)~\cite{xie2020tbg6,PhysRevB.98.081102,PhysRevB.98.075154,Potasz2021}. 

\begin{figure*}[htp!]
\includegraphics[width=0.85\textwidth]{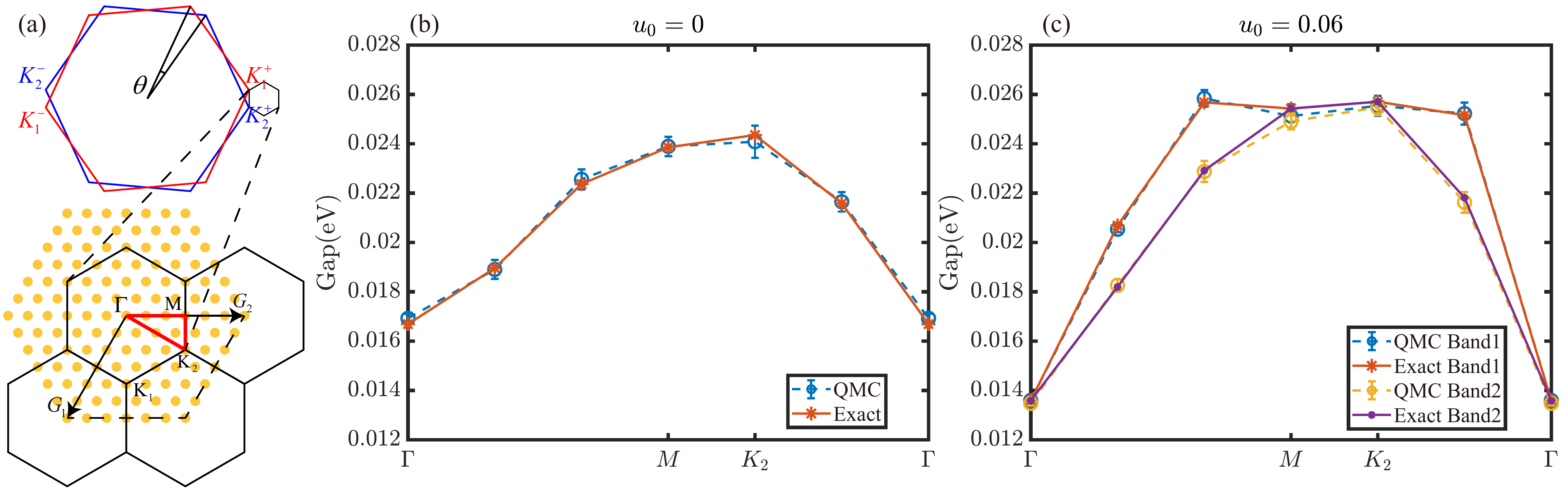}
\caption{(a) Upper panel shows the Brilliouin zone foldings in TBG, with the small hexagon representing the mBZ. $\bK^{\pm}_{1,2}$ are the Dirac points for valley $\pm$ and layer 1 and 2, $\theta$ is the rotating angle. In lower panel, $\bG_1$ and $\bG_2$ are the reciprocal lattice vectors of the mBZ. The yellow dots are the allowed momentum transfer $\bq$ points in QMC. Here we consider momentum transfer upto $\bG$, which means for a $6\times 6$ mesh in mBZ, the allowed number of $\bq$ is 126. (b) and (c) show the comparison at the CNP of the single-particle excitation gap at the chiral limit ($u_0=0$ eV) and realistic case ($u_0=0.06$ eV) between exact solution and the QMC results. The QMC gaps are obtained from fitting the imaginary time decay of the fermoin Green's function.}
	\label{fig:fig1}
\end{figure*}

{\it Continuum model}\,---\, We start from the BM model~\cite{Trambly2010,Trambly2012,Bistritzer12233,ROZHKOV20161,Santos2007,Santos2012} in plane wave basis:  $H^{\tau}_{BM} \left(\mathbf{k}\right)=\sum_{\mathbf{k {}^\prime}}H^{\tau}_{BM} {}_{\mathbf{k},\mathbf{k {}^\prime}}\; e^{-i\mathbf{k} \cdot \br}e^{i\mathbf{k \prime} \cdot \br} $ where	
$H^{\tau}_{BM} {}_{\mathbf{k},\mathbf{k {}^\prime}}=\delta_{\mathbf{k},\mathbf{k {}^\prime}}\left(\begin{array}{cc} 
				-\hbar v_F ({\bk}-\bK_1^{\tau}) \cdot \pmb{\sigma}^{\tau}  &  U_0  \\
				U_0^\dagger  & -\hbar v_F ({\bk}-\bK_2^{\tau}) \cdot \pmb{\sigma}^{\tau}
			\end{array}\right)  %\nonumber\\
			+\left(\begin{array}{cc} 
				0  &  U_1^{\tau} \delta_{\mathbf{k},\mathbf{k {}^\prime}+\tau \bG_1 }  \\ 
				U_1^{\tau \dagger} \delta_{\mathbf{k},\mathbf{k {}^\prime}-\tau \bG_1 }  & 0
			\end{array}\right) %\nonumber\\
			+\left(\begin{array}{cc} 
			0  &  U_2^{\tau} \delta_{\mathbf{k},\mathbf{k {}^\prime}+\tau(\bG_1+\bG_2) }  \\ 
			U_2^{\tau \dagger} \delta_{\mathbf{k},\mathbf{k {}^\prime}-\tau(\bG_1+\bG_2) }  & 0
		\end{array}\right),$
with $\tau=\pm$ the valley index, $\pmb{\sigma}^{\tau}=(\tau \sigma_x,\sigma_y)$ defines the A,B sublattices of the monolayer graphene. $\bK^{\tau}_1$ and $\bK^{\tau}_2$ are the corresponding Dirac points of the bottom and top layers of graphene that are now twisted by angles $\mp\frac{\theta}{2}$, and
$\bk\in$ mBZ and $\bG_1$ and $\bG_2$ are the reciprocal vectors of mBZ, as shown in Fig.~\ref{fig:fig1} (a). The interlayer tunneling between the the Dirac states is described by the matrix $U_0=
	\left(\begin{array}{cc} 
		u_0  & u_1 \\ 
		u_1 & u_0
	\end{array}\right)$, $U_1^{\tau}=   \left(\begin{array}{cc} 
		u_0  & u_1 e^{-\tau \frac{2\pi}{3}i} \\ 
		u_1 e^{\tau \frac{2\pi}{3}i} & u_0
	\end{array}\right)$ and $U_2^{\tau}= \left(\begin{array}{cc} 
		u_0  & u_1 e^{\tau \frac{2\pi}{3}i} \\ 
		u_1 e^{-\tau \frac{2\pi}{3}i} & u_0
	\end{array}\right)$
where $u_0$ and $u_1$ are the intra-sublattice and inter-sublattice interlayer tunneling amplitudes.
The flatness of the lowest two bands per spin per valley in the chiral limit ($u_0$=0) is determined by the dimensionless parameter $\alpha=\frac{u_1}{\hbar v_F k_{\theta}}$ with $k_{\theta}=8\pi\sin(\theta/2)/(3a_0)$ and the lattice constant of the monolayer graphene  $a_0$=0.246 nm.
In this paper, we choose $\hbar v_F/a_0$=2.37745 eV,
the twist angle  $\theta$=1.08$^{\circ}$ and
$u_1$=0.11 eV which leads to $\alpha$=0.586, the value corresponding to the first magic angle where the lowest two bands become completely flat in the chiral limit. We perform the QMC simulation at both the chiral limit $(u_0=0)$ and the more realistic case $(u_0=0.06$ eV), which leads to a bandwidth of 1.08 meV. 

The eigenstate of $H_{BM}^{\tau}$ can be written in the Bloch wavefunction form $\psi_{m,\tau,\bk}^X(\br)=\sum_{\bG}u_{m,\tau;\bG,X }(\bk) e^{i(\bk+\bG)\cdot \br}$
where $X=\{A_1, B_1, A_2, B_2\}$ denotes the layer and sublattice indices and $u_{m,\tau;\bG, X}(\bk)$ is the Bloch wave-function with the eigen-energy $\epsilon_{m\bk\tau}$.
Here, $m$ and $\tau$ are the band and valley indices and we omit the spin
index $s$ for now since the Hamiltonian is spin independent. The range of $m$ can be large (consider the couplings $\{ \bk,\bk+\bG_1,\bk+\bG_1+\bG_2, \cdots \}$ in Fig.~\ref{fig:fig1} (a) which has $m\in 1,2,\cdots,M$ elements, then $H^{\tau}_{BM} {}_{\mathbf{k},\mathbf{k {}^\prime}}$ is a $4M \times 4M$ matrix), we select the two flat bands and denote them as $m=1,2$ and consider the projected Coulomb interactions onto these bands in this work.

{\it Interaction model and QMC Implementations}\,---\, We develop a momentum space QMC scheme to solve the interaction Hamiltonian in the band basis:
\begin{equation}
	H_{int} = \frac{1}{2 \Omega} \sum_{\bq,\bG,|\bq+\bG|\neq0} V(\bq+\bG) \delta \rho_{\bq+\bG} \delta \rho_{-\bq-\bG},
	\label{eq:eq1}
\end{equation}
as shown in Fig.~\ref{fig:fig1} (a), $\bq\in$ mBZ and $\bq+\bG$ represents a vector in extended mBZ, and we consider the momentum transfer upto the distance of $\bG_1$ and $\bG_2$~\cite{song2020tbg2,bernevig2020tbg3}, as denoted by the yellow dots in the figure. %Here we didn't choose the normal order form of the interaction, which only adds a constant shift %$\mu \sum_{\bq,\bG} d^{\dagger}_{\bq} d_{\bq}$ 
%of energy away from the present one~\cite{bernevig2020tbg3}. 
The single gate Coulomb interaction is long-ranged: $V(\bq) =\frac{e^{2}}{4 \pi \varepsilon} \int d^{2} \br\left(\frac{1}{\br}-\frac{1}{\sqrt{\br^{2}+d^{2}}}\right) e^{i \bq \cdot \br}=\frac{e^{2}}{2 \varepsilon} \frac{1}{q}\left(1-e^{-q d}\right)$
%The general form of the long-range Coulomb interaction in plane wave basis $ c_{\bk,\alpha,s,l}$ is shown below,
%\begin{widetext}
%\begin{eqnarray}
%	H_{int} &=&\frac{1}{2 \Omega} \sum_{\bk, \bk', \bq \in GBZ} \sum_{\alpha, \alpha', s, s', l, l'} V(\bq) (c_{\bk+\bq,\alpha,s,l}^{\dagger} c_{\bk,\alpha,s,l} - \frac{1}{2} \delta_{\bq,0} )(c_{\bk'-\bq,\alpha',s',l'}^{\dagger} c_{\bk',\alpha',s',l'} - \frac{1}{2} \delta_{\bq,0})  \nonumber\\
%	\label{eq:eq12}
%\end{eqnarray}
%\end{widetext}
%where $\alpha,s,l$ represent sublattice, spin and layer index. 
%\begin{eqnarray}
%	V(\bq) &=&\frac{e^{2}}{4 \pi \varepsilon} \int d^{2} \br\left(\frac{1}{\br}-\frac{1}{\sqrt{\br^{2}+d^{2}}}\right) e^{i \bq \cdot \br}  \nonumber\\
%%	&=&\frac{e^{2}}{4 \pi \varepsilon} \int_{0}^{\infty} d r\left(\frac{1}{r}-\frac{1}{\sqrt{r^{2}+d^{2}}}\right) r \int_{0}^{2 \pi} d \theta e^{i q r \cos (\theta)}  \nonumber\\
%	&=&
%	\label{eq:eq4}
%\end{eqnarray}
with $\frac{d}{2}$ is the distance between graphene layer and single gate and $\epsilon$ is the dielectric constant. 

The definition of $\delta \rho_{\bq+\bG}$ is:
\begin{widetext}
\begin{equation}
 \delta \rho_{\bq+\bG} = \sum_{\bk \in mBZ, m_{1}, m_{2},\tau, s} \lambda_{m_1,m_2,\tau}(\bk,\bk+\bq+\bG) (d_{\bk, m_{1}, \tau, s}^{\dagger} d_{\bk+\bq, m_{2}, \tau, s} - \frac{1}{2} \delta_{\bq,0} \delta_{m_1,m_2})=(\delta\rho_{-\bq-\bG})^{\dagger}
\end{equation}
\end{widetext}
with the form factor $\lambda$ defined as
$	\lambda_{m_1 m_2, \tau}(\bk,\bk+\bq+\bG)=\sum_{\bG',X}u^{*}_{m_1,\tau;\bG',X}(\bk)u_{m_2,\tau;\bG'+\bG,X}(\bk+\bq).$ Physically, $\delta \rho$ is the electron density operator relative to the decoupled bilayer graphene at the CNP. The interaction in Eq.~\eqref{eq:eq1} differs from the normal ordered version as it can properly account for the renormalization effect of the remote bands to the flat bands~\cite{bernevig2020tbg3}.
%Here, momentum $\bq$ is defined in the extend mBZ and using the definition $d_{\bk+\bG,s,\tau}=d_{\bk,s,\tau}$ and $u_{m,\tau;\bG,X}(\bk+\bG')=u_{m,\tau;\bG+\bG',X}(\bk)$, we can write $\bq$ as $\bq=\bar{\bq}+\bG$, so that $\bar{\bq}$ is inside the mBZ,
The symmetry properties of $\lambda_{m_1,m_2,\tau}(\bk,\bk+\bq+\bG)$ are discussed in Supplemental Material (SM)~\cite{suppl}. One sees $\delta \rho_{\bq+\bG}$ in single particle basis is a block diagonal matrix according to the $\tau,s$ indices. Since $V(\bq+\bG)=V(-\bq-\bG)$, terms in $H_{int}$ can be written in a form ready to be QMC decoupled as,
\begin{widetext}
\begin{equation}
	\sum_{\bq,\bG,|\bq+\bG|\neq0} V(\bq+\bG) \delta \rho_{\bq+\bG} \delta \rho_{-\bq-\bG} = \sum_{|\bq+\bG|\neq0} \frac{V(\bq+\bG)}{2} \left[\left(\delta\rho_{-\bq-\bG}+\delta\rho_{\bq+\bG}\right)^{2}\right.-\left.\left(\delta\rho_{-\bq-\bG}-\delta\rho_{\bq+\bG}\right)^{2}\right]
	\label{eq:eq2}
\end{equation}
where $\sum_{|\bq+\bG|\neq0}$ means summation to half of the allowed values of $\bq$ and $\bG$.

According to the discrete Hubbard-Stratonovich transformation~\cite{PhysRevX.11.011014,PhysRevLett.123.157601,Assaad2008}, $e^{\alpha \hat{O}^{2}}=\frac{1}{4} \sum_{l=\pm 1,\pm 2} \gamma(l) e^{\sqrt{\alpha} \eta(l) \hat{o}}+O\left(\alpha^{4}\right)$, 
where $\gamma(\pm 1)=1+\frac{\sqrt{6}}{3}$, $\gamma(\pm 2)=1-\frac{\sqrt{6}}{3}$, $\eta(\pm 1)=\pm \sqrt{2(3-\sqrt{6})}$ and $\eta(\pm 2)=\pm \sqrt{2(3+\sqrt{6})}$, we can rewrite the partition function of Hamiltonian Eq.~\eqref{eq:eq1} in the imaginary time discretization as,

\begin{eqnarray}
&Z&=\Tr\{\prod_{t}e^{-\Delta \tau H_{i n t}(t)}\} =\Tr\{\prod_{t} e^{-\Delta \tau \frac{1}{4 \Omega}\sum_{|\bq+\bG|\neq0} V(\bq+\bG)\left[\left(\delta\rho_{-\bq-\bG}+\delta\rho_{\bq+\bG}\right)^{2}-\left(\delta\rho_{-\bq-\bG}-\delta\rho_{\bq+\bG}\right)^{2}\right]  }\} \nonumber\\
	&\approx& \sum_{\{l_{|\bq|,t}\}} \prod_{t} [ \prod_{|\bq+\bG|\neq0}\frac{1}{16} \gamma\left(l_{|\bq|_1,t}\right) \gamma\left(l_{|\bq|_2,t}\right)]  \Tr\{\prod_{t}[\prod_{|\bq+\bG|\neq0}e^{i \eta\left(l_{|\bq|_1,t}\right) A_{\bq}\left(\delta\rho_{-\bq}+\delta\rho_{\bq}\right)} e^{\eta\left(l_{|\bq|_2,t}\right) A_{\bq}\left(\delta\rho_{-\bq}-\delta\rho_{\bq}\right)}]\}
	% \nonumber\\
	%&\cdot& \sum_{l_{0,t}} \frac{1}{4} \gamma\left(l_{0,t}\right) e^{i\eta\left(l_{0,t}\right) \sqrt{2} A_{0}\delta\rho_{0}}
\label{eq:eq5}
\end{eqnarray}
\end{widetext}
where $t$ is the imaginary time index with step $\Delta\tau$, $A_{\bq+\bG} =\sqrt{\frac{\Delta \tau}{4} \frac{V(\bq+\bG)}{\Omega}}$ and $\{l_{|\bq|_1,t},l_{|\bq|_2,t},l_{0,t}\}$ are the four-component auxiliary field that lives in the space-time configuration of the path-integral. For each realization of the auxiliary field configuration, the fermion determinant can be evaluated exactly as the configurational weight, the QMC simulation is performed along a Markov chain of such configurations and the important sampling can be carried out with the physical observables (such as single-particle Green's function) computed through ensemble average~\cite{XYXu2019}. 

One shall be careful about the approximation $"\approx"$ in Eq.~\eqref{eq:eq5}. Since $\delta\rho_{\bq+\bG} \delta\rho_{\bq^{\prime}+\bG^{\prime}}-\delta\rho_{\bq^{\prime}+\bG^{\prime}} \delta\rho_{\bq+\bG}=\sum_{\bk, m_{1}, m_{2},\tau,s}[\lambda_\tau(\bk, \bk+\bq+\bG) \lambda_\tau\left(\bk+\bq,\bk+\bq+\bq^{\prime}+\bG^{\prime}\right)-\lambda_\tau\left(\bk, \bk+\bq^{\prime}+\bG^{\prime}\right) \lambda_\tau\left(\bk+\bq^{\prime}, \bk+\bq^{\prime}+\bq+\bG\right)]_{m_{1}, m_{2}} \\ d_{\bk,m_1,\tau,s}^{\dagger} d_{\bk+\bq+\bq^{\prime},m_2,\tau,s}$, which means $\left[\delta\rho_{\bq+\bG}, \delta\rho_{\bq^{\prime}+\bG^{\prime}}\right] \neq 0$. However when the number of $\bq$ is limited as shown in Fig.~\ref{fig:fig1} (a), i.e. allowing momentum transfer upto $\bG$, our results show the systematic discretization errors are acceptable. By setting $\varepsilon=7 \varepsilon_{0}$, the moir\'e lattice vector $L_{M}=\frac{a_{0}}{2 \sin \left(\frac{\theta}{2}\right)}$, the area of system $\Omega=N_{\bk}\frac{\sqrt{3}}{2} L_{M}^{2}$ with $N_{\bk}$ the number of $\bk$ points in mBZ (here we have $N_{\bk}=36$ for the $6\times6$ mesh), and the gate distance $d=40$ nm, mBZ reciprocal lattice vector $|\bG|=\frac{4 \pi}{\sqrt{3} L_{M}}$,  we have
$	\frac{V(\bar{\bq})}{\Omega} \approx 0.01585 \frac{1}{\sqrt{N_{k}}\bar{\bq}}\left(1-e^{-22.36 \frac{1}{\sqrt{N_{k}}} \bar{\bq}}\right)$ eV where $\bar{\bq}$ is the distance between momenta in mBZ by setting two nearest $\bk$ points with unit length.

{\it Discussion of the sign-problem}\,---\, We note in Eq.~\eqref{eq:eq5}, the exponential parts of decoupled interaction are anti-Hermitian, the following three statements about the sign structure of the QMC fermion determinants %for the cases of single valley and single spin, single valley and double spin, single spin and double valley, 
are in order:

\textbf{Statement 1}
Considering single valley and single spin without kinetic terms at half filling, the sign of the determinant is always real.

\textbf{Proof}
In our decoupled Hamiltonian, the configurational probability is proportional to
$e^{-\frac{1}{2}\sum_{j}\operatorname{Tr}(M_{j})}\operatorname{det}\left(I+e^{M_{1}}e^{M_{2}}...e^{M_{n}}\right)$. Here $M_{j}$ are anti-Hermitian matrices in single particle basis and $e^{-\frac{1}{2}\sum_{j}\operatorname{Tr}(M_{j})}$ comes from constant terms in $\delta\rho_{\bq+\bG}$. Since $e^{M_j}$ are unitary matrices, $U=e^{M_{1}}e^{M_{2}}...e^{M_{n}}$ is also unitary with eigenvalue $e^{i\lambda_j}$. Set $\operatorname{det}\left(U\right)=e^{\sum_{j}\operatorname{Tr}(M_{j})}=e^{\sum_{j}i\lambda_j}=e^{i\Gamma}$, $e^{-i\frac{\Gamma}{2}}\operatorname{det}\left(I+U\right)=e^{-i\frac{\Gamma}{2}}\prod_{j}\left(1+e^{i \lambda_{j}}\right) .$ For any term $e^{i (\Sigma_{k\in A} \lambda_{k} -\frac{\Gamma}{2})}$, $A\subseteq \left\lbrace 1,2,...,n\right\rbrace $, there is always a term $e^{i (\Sigma_{k\notin A} \lambda_{k}-\frac{\Gamma}{2})}=e^{-i (\Sigma_{k\in A} \lambda_{k}-\frac{\Gamma}{2})}$, so add all terms together will always be real. 

%In the mean time, if the largest $\left|\lambda_{j}\right|$ is less than $\frac{\pi}{N_{\bk}}$ ($N_{\bk}$ is matrix dimension of $M_j$), then at high temperature (small $\beta$) all terms are positive rendering $e^{-i\frac{\Gamma}{2}}\operatorname{det}\left(I+U\right)$ positive.

\textbf{Statement 2}
Considering single valley and double spin without kinetic terms at half filling, there is no sign problem.

\textbf{Proof}
It is straightforward to see this result according to Statement 1 by noticing the other spin just gives a copy so that the real sign will become a non-negative sign. But kinetic terms could change this result. See SM~\cite{suppl} for details.

\textbf{Statement 3}
Considering single spin and double valley with flat band kinetic terms at half filling, there is no sign problem.

\textbf{Proof}
One can relabel $d_{k,m,-\tau,s}$ as $-m* \tilde{d}_{k,-m,-\tau,s}^{\dagger}$ in $-\tau$ subspace, then prove the single-particle matrixes between two valleys satisfy
$\delta \rho_{\bq+\bG,-\tau} = -\delta \rho_{-\bq-\bG,\tau}$. Thus $M_{j,-\tau}=M_{j,\tau}^*$. And this transformation will keep flat band kinetic matrices intact between two valleys. So the determinant of valley $-\tau$ is complex conjugated with that of valley $\tau$. We note similar observation, that the TBG Hamiltonian at CNP after QMC decoupling is invariant under anti-unitary particle-hole symmetry and thus free of the sign problem, has also been pointed out in Ref.~\cite{JYLee2021}. %See Appendix \ref{app:app1} for details.

We organize these statements in Tab.~\ref{tab1}. And we noted that Ref.~\cite{Hofmann2021} also shows some similar results.
\begin{table}[htp!]
	\caption{List of the sign structure for TBG Hamiltonian.}
	\begin{ruledtabular}
		\begin{tabular}{cccc}
			\textrm{Degrees of freedom}&
			\textrm{Kinetic terms}&
			\textrm{Sign Structure}\\
			\colrule
			Single valley single spin & No & Real \\
			Single valley double spin & No & Non-negative \\
			Double valley single spin & Flat bands & Non-negative \\
			Double valley double spin & Flat bands & Non-negative
		\end{tabular}
	\end{ruledtabular}
\label{tab1}
\end{table}

\begin{figure}[htp!]
\includegraphics[width=0.9\columnwidth]{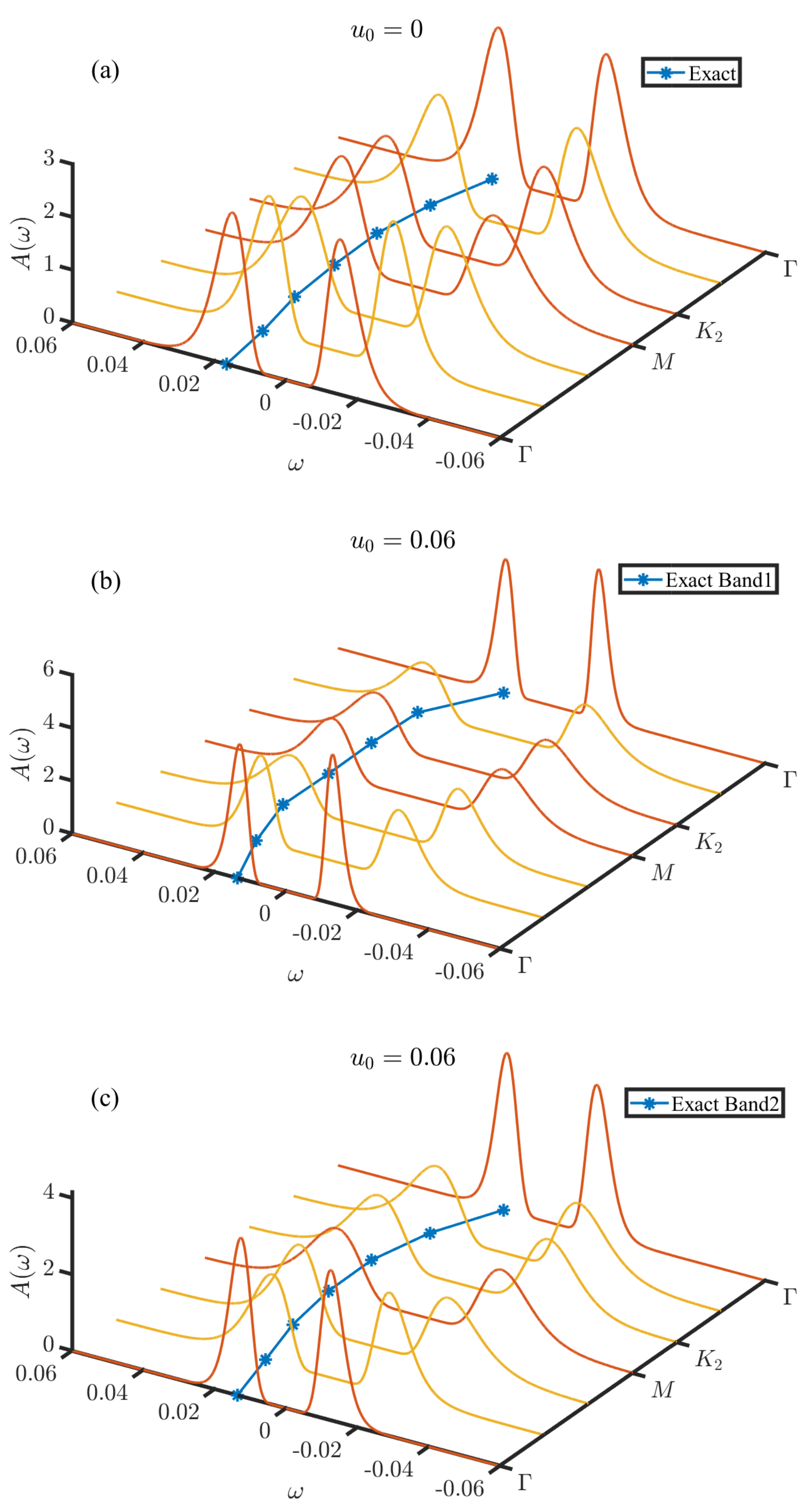}
\caption{Single-particle spectra obtained from QMC+SAC at CNP with (a) chiral limit ($u_0=0$) and (b), (c) $u_0=0.06$. The exact gaps are the same as in Fig.~\ref{fig:fig1} (b) and (c).}
\label{fig:fig2}
\end{figure}

{\it QMC Results and Discussions}\,---\, With such understanding, we carried out QMC simulations at CNP for TBG Hamiltonian in momentum-space, on the grids of $6\times 6$ in mBZ, both at chiral limit $(u_0=0)$ and with realistic parameter $(u_0=0.06)$. To make sure our QMC have converged to the ground state, we set the temperature $T=0.667$ meV in the simulations, which turns out to be magnitudes smaller than the obtained gap, and divide the inverse temperature $\beta=1/T$ to 150 pieces with $\Delta\tau=0.01$ such that the Trotter error negligible.

Fig.~\ref{fig:fig1} (b) and (c) show the comparison of the single-particle excitation gap at the chiral limit ($u_0=0$) and $u_0=0.06$, obtained from fitting the imaginary decay of the Green's function in QMC simulations, respectively. At the chiral limit, the two bands are degenerate, whereas in the realistic case, we diagonalize the Green's function at every $\bk$ point in the $2\times2$ band basis. The agreement between the exact gaps~\cite{bernevig2020tbg5} and the QMC ones is perfect.

Fig.~\ref{fig:fig2} show the single-particle spectra, obtained from applying
stochastically analytic continuation (SAC)~\cite{PhysRevB.57.10287,beach2004identifying,PhysRevE.94.063308,PhysRevB.78.174429,PhysRevX.7.041072} upon the imaginary time Green’s function from QMC simulations. Such QMC+SAC scheme has been shown to reliably reveal many interesting dynamical features in various strongly correlated systems~\cite{PhysRevX.7.041072,PhysRevLett.121.077201,PhysRevB.98.174421,PhysRevLett.120.167202,yan2021topological,li2020kosterlitz,hu2020evidence,zhou2020amplitude,PhysRevB.103.014408,jiang2020}. Fig.~\ref{fig:fig2} (a) shows the single-particle spectrum at the chiral limit ($u_0=0$) where the two bands are degenerate and Fig.~\ref{fig:fig2} (b) and (c) the spectral of the two bands at $u_0=0.06$ where they are not degenerate. In all cases, the spectral are particle-hole symmetric.

%Fig.~\ref{fig:fig2} shows the excitation gap at $\nu=3/4$ with realistic parameters, at this filling, the ground state have been revealed from other works to be a QAH with the present parameters~\cite{BBChen2020,YiZhang2020,JPLiu2021}. Fig.~\ref{fig:fig2} (a) show the comparison of the single-particle gap between QMC and Hartree-Fock mean-field~\cite{YiZhang2020}, it is interesting to see that the QMC gaps are slightly larger than the mean-field ones, signifying the non-perturbative interaction effects. Fig.~\ref{fig:fig2} (b) shows the QMC Berry curvature for such gapped state. In the computation, we add a small band-dependent hopping term to lift the band degeneracy, and the procedure is described in detail in Appendix~\ref{app:app2}. It is obvious that the non-trivial Berry curvatures are accumulated close the $\Gamma$ point in mBZ and after integrating over the mBZ, we obtain the Chern number $C=1$.

%-- the fermion determinants of the interaction Hamiltonian Eq.~\eqref{eq:eq1} are without sign-problem even if the kinetic terms are included for both valleys -- 
With the establishment of such unbiased QMC computational framework, as summarized in Tab.~\ref{tab1}, the controlled computation pathway for the interaction effects in realistic TBG model is clearly opening up. The questions of the nature of the ground states at different integer fillings and the TBG phase diagram at different twist angle, chiral ratio, hNB alignment, the skyrmion SC and the dynamic and spectral properties, etc, can now be investigated as those have been investigated with QMC simulations in other exotic strongly correlated electron systems~\cite{XYXu2020,WWang2021,GPPan2021,ChuangChen2021}.

{\it Acknowledgments}\,---\, 
ZYM thanks Xi Dai for the insightful discussion and continuous encouragement for addressing the momentum-space solutions of TBG. XZ, GPP and ZYM acknowledge support from the RGC of Hong Kong SAR of China (Grant Nos. 17303019, 17301420 and AoE/P-701/20), MOST through the National Key Research and Development Program (Grant No. 2016YFA0300502) and the Strategic Priority Research Program of the Chinese Academy of Sciences (Grant No. XDB33000000). YZ is supported in part by the strategic priority research program of the Chinese Academy of Sciences Grant No. XDB28000000 and NSFC Grant Nos. 11674278, 12004383 and 12074276 and the Fundamental Research Funds for the Central Universities. JK acknowledges the support from the NSFC Grant No.~12074276, and the Priority Academic Program Development (PAPD) of Jiangsu Higher Education Institutions. We thank the Computational Initiative at the Faculty of Science and the Information Technology Services at the University of Hong Kong for their technical support and generous allocation of CPU time.
% Comment out the next line to display the title of the article	
\bibliographystyle{apsrev4-1}

\bibliography{TBG.bib}

\clearpage
\onecolumngrid
%\appendix
\begin{center}
	\textbf{Supplemental Material for "Momentum space quantum Monte Carlo on twisted bilayer Graphene"}
\end{center}

\section{Properties of the form factor $\lambda$ and proof of the sign structure}
		\label{app:app1}
		Here we follow the discussion in band basis as in Ref.\cite{bernevig2020tbg3}.
		
		The Hermiticity condition for $\lambda$ is
		\begin{equation}
		\lambda_{m,n,\tau}(\bk,\bk+\bq+\bG)=\lambda_{n,m,\tau}^{*}(\bk+\bq+\bG,\bk).
		\end{equation}
		
		The $C_{2z}T$ symmetry operates on $\lambda$ as
		\begin{equation}
		\lambda_{m,n,\tau}(\bk,\bk+\bq+\bG)=\lambda_{m,n,\tau}^{*}(\bk,\bk+\bq+\bG),
		\end{equation}
		which means $\lambda$ is real.
		
		The $C_{2z}P$ symmetry operates on $\lambda$ as
		\begin{equation}
		\lambda_{m,n,\tau}(\bk,\bk+\bq+\bG)=m*n*\lambda_{-m,-n,-\tau}(\bk,\bk+\bq+\bG).
		\end{equation}
		
		Based on these relations, we can derive the Statement 3 in the main text as:
		\begin{eqnarray}
		\delta \rho_{\bq+\bG,-\tau} &=& \sum_{\bk, m,n} \lambda_{m,n,-\tau}(\bk,\bk+\bq+\bG) (d_{\bk,m,-\tau}^{\dagger} d_{\bk+\bq,n, -\tau} - \frac{1}{2} \delta_{\bq,0} \delta_{m,n}) \nonumber\\
		&=& \sum_{\bk, m,n} -m*n*\lambda_{m,n,\tau}(\bk,\bk+\bq+\bG) (d_{\bk+\bq,-n, -\tau} d_{\bk,-m,-\tau}^{\dagger} - \frac{1}{2} \delta_{\bq,0} \delta_{m,n}) \nonumber\\
		&=& \sum_{\bk, m,n} -\lambda_{m,n,\tau}(\bk,\bk+\bq+\bG) (\tilde{d}_{\bk+\bq,n, -\tau}^{\dagger} \tilde{d}_{\bk,m,-\tau} - \frac{1}{2} \delta_{\bq,0} \delta_{m,n}) \nonumber\\
		&=& \sum_{\bk, m,n} -\lambda_{n,m,\tau}(\bk,\bk-\bq-\bG) (\tilde{d}_{\bk,n, -\tau}^{\dagger} \tilde{d}_{\bk-\bq,m,-\tau} - \frac{1}{2} \delta_{\bq,0} \delta_{m,n})
		\end{eqnarray}
		Here $\tilde{d}_{\bk,m,-\tau}=m*d_{\bk,-m,-\tau}^{\dagger}$, $m,n\in{\left\lbrace \pm1 \right\rbrace}$. According to the last line, single-particle matrices in the fermion determinant between two valleys satisfy
		$\delta \rho_{\bq+\bG,-\tau} = -\delta \rho_{-\bq-\bG,\tau}$.
		Besides, this transformation does not change kinetic terms. Since
		\begin{equation}
		\varepsilon_{m,\tau}(\bk)=-\varepsilon_{-m,\tau}(-\bk)=\varepsilon_{m,-\tau}(-\bk)
		\end{equation}
		So that
		\begin{eqnarray}
		&&\varepsilon_{m,\tau}(\bk)d_{\bk,m,\tau}^{\dagger}d_{\bk,m,\tau}+\varepsilon_{-m,\tau}(-\bk)d_{-\bk,-m,\tau}^{\dagger}d_{-\bk,-m,\tau}	\nonumber\\
		&=&\varepsilon_{m,\tau}(\bk)(d_{\bk,m,\tau}^{\dagger}d_{\bk,m,\tau}-d_{-\bk,-m,\tau}^{\dagger}d_{-\bk,-m,\tau})
		\end{eqnarray}
		\begin{eqnarray}
		&&\varepsilon_{-m,-\tau}(\bk)d_{\bk,-m,-\tau}^{\dagger}d_{\bk,-m,-\tau}+\varepsilon_{m,-\tau}(-\bk)d_{-\bk,m,-\tau}^{\dagger}d_{-\bk,m,-\tau}	\nonumber\\
		&=&\varepsilon_{m,\tau}(\bk)(d_{\bk,-m,-\tau}d_{\bk,-m,-\tau}^{\dagger}-d_{-\bk,m,-\tau}d_{-\bk,m,-\tau}^{\dagger})	\nonumber\\
		&=&\varepsilon_{m,\tau}(\bk)(\tilde{d}_{\bk,m,-\tau}^{\dagger}\tilde{d}_{\bk,m,-\tau}-\tilde{d}_{-\bk,-m,-\tau}^{\dagger}\tilde{d}_{-\bk,-m,-\tau})
		\end{eqnarray}
		One can see kinetic terms can be viewed as complex conjugated between two valleys since the dispersion $\varepsilon_{m,\tau}(\bk)$ is real. We note similar observation, that the TBG Hamiltonian at CNP after QMC decoupling is invariant under anti-unitary particle-hole symmetry, is also pointed out in Refs.~\cite{JYLee2021,Hofmann2021}. 
		
		We derive the Statement 2 in another way, different from that in the main text, and then discuss the flat band kinetic terms in this case.
		\begin{eqnarray}
		\delta \rho_{\bq+\bG,-s} &=& \sum_{\bk, m,n} \lambda_{m,n}(\bk,\bk+\bq+\bG) (d_{\bk,m,-s}^{\dagger} d_{\bk+\bq,n,-s} - \frac{1}{2} \delta_{\bq,0} \delta_{m,n}) \nonumber\\
		&=& \sum_{\bk, m,n} -\lambda_{n,m}(\bk,\bk-\bq-\bG) (d_{\bk,n,-s}d_{\bk-\bq,m,-s}^{\dagger} - \frac{1}{2} \delta_{\bq,0} \delta_{m,n}) \nonumber\\
		&=& \sum_{\bk, m,n} -\lambda_{n,m}(\bk,\bk-\bq-\bG) (\tilde{d}_{\bk,n,-s}^{\dagger}\tilde{d}_{\bk-\bq,m,-s} - \frac{1}{2} \delta_{\bq,0} \delta_{m,n}) \nonumber\\
		\end{eqnarray}
		In the last line we just define $\tilde{d}_{\bk,m,-s}=d_{\bk,m,-s}^{\dagger}$, single-particle matrices in the fermion determinant between two spins also satisfy $\delta \rho_{\bq+\bG,-s} = -\delta \rho_{-\bq-\bG,s}$. But this transformation add a minus sign to the kinetic terms.
		\begin{eqnarray}
		&&\varepsilon_{m}(\bk)d_{\bk,m,-s}^{\dagger}d_{\bk,m,-s}+\varepsilon_{-m}(-\bk)d_{-\bk,-m,-s}^{\dagger}d_{-\bk,-m,-s}	\nonumber\\
		&=& -\varepsilon_{m}(\bk)(d_{\bk,m,-s} d_{\bk,m,-s}^{\dagger} - d_{-\bk,-m,-s} d_{-\bk,-m,-s}^{\dagger})	\nonumber\\
		&=& -\varepsilon_{m}(\bk)(\tilde{d}_{\bk,m,-s}^{\dagger} \tilde{d}_{\bk,m,-s} - \tilde{d}_{-\bk,-m,-s}^{\dagger} \tilde{d}_{-\bk,-m,-s})
		\end{eqnarray}
		It is thus not obvious when consider both spin degree of freedom for a single valley, when including the kinetic energy, the QMC simulation is still absent of sign-problem.
		
		\section{Measurement of the Green's function in QMC}
		Here we introduce the measurement of Green's funtions in QMC. Generally, average of any observables $\hat{O}$ can be written as,
		\begin{equation}
		\left\langle \hat{O} \right\rangle = \frac{\Tr(\hat{O} e^{-\beta H})}{\Tr(e^{-\beta H})} = \sum_{\{l_{|\bq|,t}\}} \frac{P(\{l_{|\bq|,t}\}) \Tr[\prod_{t}\hat{B}_t(\{l_{|\bq|,t}\})] \frac{\Tr[\hat{O} \prod_{t}\hat{B}_t(\{l_{|\bq|,t}\})]}{\Tr[\prod_{t}\hat{B}_t(\{l_{|\bq|,t}\})]}}{\sum_{\{l_{|\bq|,t}\}} P(\{l_{|\bq|,t}\}) \Tr[\prod_{t}\hat{B}_t(\{l_{|\bq|,t}\})]} 
		\end{equation}
		According to Eq. (4) in the main text, $P(\{l_{|\bq|,t}\})=\prod_{t} [ \prod_{|\bq+\bG|\neq0}\frac{1}{16} \gamma\left(l_{|\bq|_1,t}\right) \gamma\left(l_{|\bq|_2,t}\right)]$ and 
		\begin{equation}
		\hat{B}_t(\{l_{|\bq|,t}\})=\prod_{|\bq+\bG|\neq0}e^{i \eta\left(l_{|\bq|_1,t}\right) A_{\bq}\left(\delta\rho_{-\bq}+\delta\rho_{\bq}\right)} e^{\eta\left(l_{|\bq|_2,t}\right) A_{\bq}\left(\delta\rho_{-\bq}-\delta\rho_{\bq}\right)},
		\end{equation} 
		respectively.
		
		Since $M_j$ is now of the form of fermion bilinear, one can trace out the fermion operator to obtain the determinant
		\begin{equation}
		\Tr[e^{M_{1}}e^{M_{2}}...e^{M_{n}}]=\det[I+e^{M_{1}}e^{M_{2}}...e^{M_{n}}],
		\end{equation}
		the obtained determinant is the single particle fermion determinant for QMC. Exact constant terms in $\hat{B}(\{l_{|\bq|,t}\})$ as $e^{-\frac{1}{2}\sum_{j}\operatorname{Tr}(M_{j})}$ represent the left part in the single particle basis $e^{M_{t}}$, and it is from such structure
		\begin{equation}
		e^{-\frac{1}{2}\sum_{j}\operatorname{Tr}(M_{j})}\operatorname{det}\left(I+e^{M_{1}}e^{M_{2}}...e^{M_{n}}\right)
		\end{equation} 
		we prove the Statement 1 in the main text.
		
		Now consider operator $\hat{O}=d_i(\tau) d_j^{\dagger}(\tau)$, which is the single-particle Green's function $G(\tau,\tau)_{i,j}=\langle d_i(\tau) d_j^{\dagger}(\tau)\rangle$ in QMC,
		\begin{equation}
		\frac{\Tr[\prod_{t_1>\tau}\hat{B}_{t_1}(\{l_{|\bq|,t}\}) d_i(\tau) d_j^{\dagger}(\tau) \prod_{t_2\leqslant\tau}\hat{B}_{t_2}(\{l_{|\bq|,t}\})]}{\Tr[\prod_{t}\hat{B}_t(\{l_{|\bq|,t}\})]}=\left(I+e^{M_{\tau}}e^{M_{\tau-1}}...e^{M_{1}}\cdot e^{M_{n}}e^{M_{n-1}}...e^{M_{\tau+1}}\right)_{i,j}^{-1}.
		\end{equation}
		
		\textbf{Statement 4}
		If there is no kinetic term, Green's function $G(\tau,\tau)_{i,j}$ should always be diagonal with all non-zero elements $0.5$.
		
		\textbf{Proof}
		For a certain configuration $U_S=B_{1,s_1}B_{2,s_2}...B_{n,s_n}$, ignoring matrix commutation in each $B_{t,s_1}$, we can always find another configuration  $U_{S'}=B_{1,-s_n}...B_{n-1,-s_2}B_{n,-s_1}=U_S^{-1}$ with the same weight
		\begin{equation}
		e^{-\frac{1}{2}\sum_{j}\operatorname{Tr}(M_{j})}\operatorname{det}\left(I+e^{M_{1}}e^{M_{2}}...e^{M_{n}}\right)=e^{\frac{1}{2}\sum_{j}\operatorname{Tr}(M_{j})}\operatorname{det}\left(I+(e^{M_{1}}e^{M_{2}}...e^{M_{n}})^{-1}\right).
		\end{equation} 
		We add Green's funstion $G(\tau,\tau)_{i,j}$ of this pair configurations together, which is $(I+e^{M_{1}}e^{M_{2}}...e^{M_{n}})^{-1}+(I+(e^{M_{1}}e^{M_{2}}...e^{M_{n}})^{-1})^{-1}=I$. Remeber we need to divide it by 2 for average of these two configurations, therefore after sampling and average, $G(\tau,\tau)_{i,j}$ should always be diagonal with all non-zero elements 0.5.
		
		Besides, consider $G(\tau,0)_{i,j}=\langle d_i(\tau) d_j^{\dagger}(0)\rangle$,
		\begin{equation}
		\frac{\Tr[\prod_{t_1>\tau}\hat{B}_{t_1}(\{l_{|\bq|,t}\}) d_i(\tau) \prod_{t_2\leqslant\tau}\hat{B}_{t_2}(\{l_{|\bq|,t}\})d_j^{\dagger}(0)]}{\Tr[\prod_{t}\hat{B}_t(\{l_{|\bq|,t}\})]}=(G(\tau,\tau)\cdot e^{M_{\tau}}e^{M_{\tau-1}}...e^{M_{1}})_{i,j}
		\end{equation}
		
		We also introduce a useful way below to obtain real valued measurables for other fillings, in which, the sign of the weight is always real. The definition of $\delta \rho_{\bq+\bG}$ should be rewritten as:
		\begin{equation}
		\delta \rho_{\bq+\bG} = \sum_{\bk \in mBZ, m_{1}, m_{2},\tau, s} \lambda_{m_1,m_2,\tau}(\bk,\bk+\bq+\bG) (d_{\bk, m_{1}, \tau, s}^{\dagger} d_{\bk+\bq, m_{2}, \tau, s} - C \delta_{\bq,0} \delta_{m_1,m_2})=(\delta\rho_{-\bq-\bG})^{\dagger}.
		\end{equation}
		Here $C$ can be any real number.
		
		Assume we would like to measure $\langle F \rangle=\langle F^\ast \rangle= \frac{\langle F+F^\ast\rangle}{2}$. For any configuration $S$, we can define a dual configuration $-S$. For $-S$, in $\hat{B}_t(\{l_{|\bq|,t}\})=\prod_{|\bq+\bG|\neq0}e^{i \eta\left(l_{|\bq|_1,t}\right) A_{\bq}\left(\delta\rho_{-\bq}+\delta\rho_{\bq}\right)} e^{\eta\left(l_{|\bq|_2,t}\right) A_{\bq}\left(\delta\rho_{-\bq}-\delta\rho_{\bq}\right)}$, all $\eta\left(l_{|\bq|_2,t}\right)$ do not change and all other $\eta\left(l_{|\bq|,t}\right)$ change to $-\eta\left(l_{|\bq|,t}\right)$. Then one can see $\hat{B}_t(\{l_{|\bq|,t}\})$ change to $\hat{B}_t^\ast(\{l_{|\bq|,t}\})$, weight $P_S= e^{-C\sum_{j}\operatorname{Tr}(M_{j})}\operatorname{det}\left(I+e^{M_{1}}e^{M_{2}}...e^{M_{n}}\right)$ changes to $P_{-S}=\left[ e^{-C\sum_{j}\operatorname{Tr}(M_{j})}\operatorname{det}\left(I+e^{M_{1}}e^{M_{2}}...e^{M_{n}}\right)\right] ^\ast=P_S^\ast$. It is obvious that
		\begin{equation}
		\frac{\langle F+F^\ast\rangle}{2}=\sum_{\left\lbrace S\right\rbrace } \frac{P_S F_S+P_{-S} F^\ast_{-S}}{2}=\sum_{\left\lbrace S\right\rbrace } Re(P_S)F_S.
		\end{equation}
		We conclude here, to measure real measurables for any filling, one just need to see real part of weight as an effective weight. 
		
		\section{Stochastic Analytic Continuation (SAC) Method}
		Here we use the stochastic analytic continuation (SAC) method~\cite{PhysRevB.57.10287,beach2004identifying,PhysRevE.94.063308,PhysRevB.78.174429,PhysRevX.7.041072} to obtain the spectral function $A_{\vec{k}}(\omega)$, where $A_{\vec{k}}(\omega)=-(1 / \pi) \operatorname{Im} G_{\vec{k}}^{r e t}(\omega)$. The relationship of  $A_{\vec{k}}(\omega)$ and correlation function in imaginary time is:
		
		\begin{equation}
		G_{\vec{k}}(\tau)=\int_{-\infty}^{\infty} d \omega\left[\frac{e^{-\omega \tau}}{1+e^{-\beta \omega}}\right] A_{\vec{k}}(\omega)
		\end{equation}
		
		The following quantity $\chi^2$ is used to evaluate the quality of the fit, where we first give an very generic variational ansatz of the spectrum $A(\omega)$ and then carry out the above Laplacian transfermation and use the obtained Green's function to compare (fit) with the QMC Green's function and optimize the ansatz accroding to the $\chi^2$ stochastically:
		
		\begin{equation}
		\chi^{2}=\sum_{i j}\left(\bar{G}\left(\tau_{i}\right)-\int_{-\infty}^{\infty} d \omega\left[\frac{e^{-\omega \tau_i}}{1+e^{-\beta \omega}}\right] A(\omega)\right)\left(C^{-1}\right)_{i j}\left(\bar{G}\left(\tau_{j}\right)-\int_{-\infty}^{\infty} d \omega\left[\frac{e^{-\omega \tau_j}}{1+e^{-\beta \omega}}\right] A(\omega)\right)\end{equation}
		
		where\begin{equation} C_{i j}=\frac{1}{N_{b}\left(N_{b}-1\right)} \sum_{b=1}^{N_b}\left(G^{b}\left(\tau_{i}\right)-\bar{G}\left(\tau_{i}\right)\right)\left(G^{b}\left(\tau_{j}\right)-\bar{G}\left(\tau_{j}\right)\right)\end{equation}
		
		Here $\bar{G}\left(\tau_{i}\right)$ is the Monte Calro average of Green's functions of $N_b$ bins.
		
		In order to obtain the optimized spectral function, we perform the Monte Carlo sampling~\cite{PhysRevE.94.063308,PhysRevB.78.174429} to improve it with the QMC Green's function. The ansatz of the spectral function is $A(\omega)=\sum_{i=1}^{N_{\omega}} A_{i} \delta\left(\omega-\omega_{i}\right)$ and the weight of such Monte Carlo configuration is:
		
		\begin{equation}
		W \sim \exp \left(-\frac{\chi^{2}}{2 \,\Theta_T}\right)
		\end{equation}
		
		Here $\Theta_T$ is an analogy to temperature. We carry out simulated annealing method and at different $\Theta_T$ we compute the average $\langle\chi^{2}\rangle$, finally choose the converged $\Theta_T$ to satisfy:
		\begin{equation}
		\langle\chi^{2}\rangle=\chi_{\min }^{2}+a \sqrt{\chi_{\min }^{2}}
		\end{equation}
		usually we set $a=2$, and it is from such optimized $\Theta$ and $\chi^2$, we further compute the ensemble average of the spectra as the final ones to present in the main text.

\end{document}